\documentclass[
superscriptaddress,
 amsmath,amssymb,
 aps,
 prf,
]{revtex4-2}

\usepackage{graphicx}
\usepackage{dcolumn}
\usepackage{bm}
\usepackage{xcolor}

\begin{document}


\title{Reorientation dynamics of microswimmers at fluid-fluid interfaces}

\author{Harinadha Gidituri, Zaiyi Shen, Alois W{\"u}rger, Juho S. Lintuvuori}

\affiliation{Univ. Bordeaux $\&$ CNRS, LOMA (UMR 5798), 33405 Talence, France}

\date{\today}






\date{Received: date / Accepted: date}

\begin{abstract}
We study the orientational and translational dynamics of spherical microswimmers trapped at fluid interfaces, in terms of the force dipole and source dipole components of their flow field. Using numerical simulations and analytical calculations, we show that the force dipole exerts a torque, orienting  pushers parallel to the interface, and pullers in normal direction. The source dipole results in particle rotation only for a finite viscosity contrast between the two fluids, in agreement with previous studies. The superposition of these two contributions leads to an rotational dynamics with a steady-state orientation that depends on the relative magnitudes of the force and source dipoles. In the general case, swimmers with weak force dipoles and strong pullers are observed to align perpendicular to the interface and become stationary, while strong pushers have a finite inclination angle towards the lower viscosity fluid and swim along the interface.
\end{abstract}

\maketitle

{\it Introduction.--}
 Microscopic active particles, such as bacteria (eg. E. coli) and algae (eg. Chlamydomonas) often found in confined environments near solid-fluid and fluid-fluid interfaces \cite{Lauga_RPP_2009}, and surface interactions can affect their dynamics. For example, {\it E. coli} swims in clockwise circular motion near a no-slip surface \cite{Lauga_BPJ_2006} and anti-clockwise close to air-liquid interface \cite{Leonardo_PRL_2011}.  In the last two decades, 
 researchers have synthesized artificial microswimmers (eg. active Janus particles)~\cite{ebbens2016active}. These artificial swimmers have potential for applications, such as drug delivery \cite{Jinxing_Acsnano_2016} and environmental remediation \cite{Soler_ACSnano_2013,Wei_Nanoscale_2013,Jurado_Small_2015,Linlin_AAMI_2019}. For example, active Janus particles have been used to collect oil droplets using sea water as a fuel~\cite{Wei_Nanoscale_2013}. 
 Understanding the dynamics of these particles at interfaces, is crucial for designing such applications.

 The dynamics of microwsimmers in the vicinity of an interface is governed by the hydrodynamic stresses created by the particles~\cite{Lauga_PRL2_2008,Lauga_PoF_2014}, and has been studied both experimentally~\cite{Leonardo_PRL_2011,Ardekani_PRE_2019,Morse_BPJ_2013,Deng_Langmuir_2020,Stebe_ACIS_2017,Hollenbeck_BPJ_2014,Morikawa_JBB_2006,Angelini_PNAS_2009,Gonzalez_PRF_2020} and theoretically \cite{Pimponi_JFM_2016,Crowdy_JFM_2011, Ardekani_JFM_2017,Gidituri_SM_2019,Ardekani_JFM_2017,Ishikawa_APL_2019}. Previous investigations mainly considered microswimmers near interfaces. Less is known of particles straddling a clean interface between two fluids, where they can be thermodynamically trapped due to a Pickering effect~\cite{Pickering_CST_1907}. 
The hydrodynamic boundary condition at the contact line between the two fluids and the particle surface, can play a role both in the linear and orientational dynamics of the swimmer~\cite{Nicholas_JFM_2021}.
Simulations of self-diffusiophoretic colloids at fluid-fluid interface predicted an emergence of an aligning torque on a particle at an interface between two fluids with equal viscosities~\cite{Malgaretti_SM_2020}. 

In general, the two fluids have different viscosities, characterised by the ratio $\lambda=\eta_2/\eta_1$. The effects of viscosity stratification on the dynamics (translational and rotational) of microswimmers  have been studied both experimentally and numerically~\cite{DANIELS_microbiology_1980,Takabe_microbiology_2017,Liebchen_PRL_2018,Datt_PRL_2020,Eastham_PRF_2020,Simone_SR_2021,Stehnach_Biorxiv_2020,lopez_arxiv_2020}, where  reorientation towards negative gradients (towards the less viscous fluid) is typically observed in fluids with the viscosity changing over a length scale considerably larger than the particle size. This negative viscotaxis has also been predicted by simulations of catalytic swimmers~\cite{Malgaretti_SM_2016} and observed in recent experiments of bacteria~\cite{Simone_SR_2021,Stehnach_Biorxiv_2020} in sharp viscosity gradients.

\begin{figure}[b]
\includegraphics[width=0.5\columnwidth]{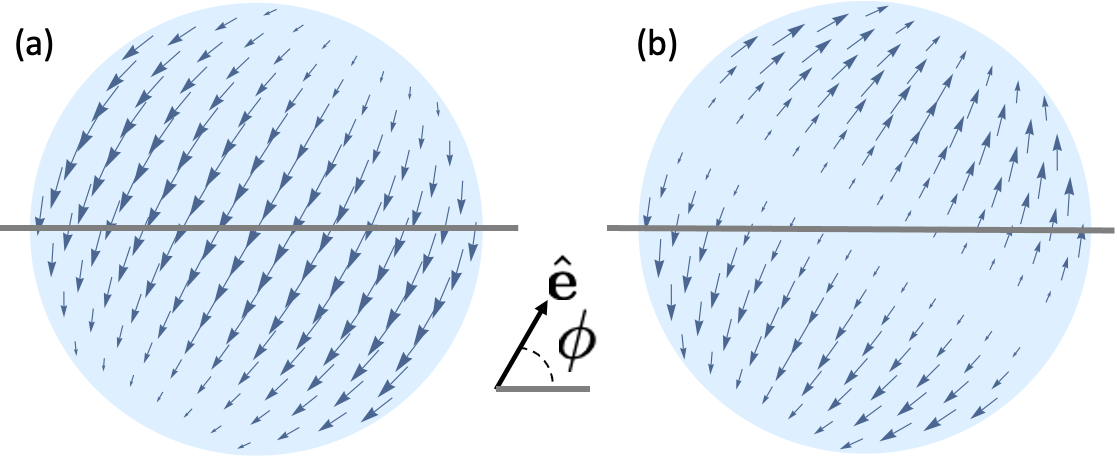}
\caption{(a,b) Slip flows at the particle surface corresponding to (a) source and (b) force dipoles, inclined at an angle $\phi$ with respect to the interface.}
\label{fig:dipoles}
\end{figure}

In this work, we use lattice Boltzmann simulations to study the dynamics of a spherical swimmer trapped at a clean interface separating two fluids. We consider the experimentally most relevant case of low Reynolds and capillary numbers, where inertial effects and interface deformation are small. Both the viscosity contrast and the hydrodynamic boundary condition are described in terms of a Ginzburg-Landau-functional for an interface of  finite thickness. The swimmer is described by the squirmer model \cite{Lighthill_CPAM_1952}, the flow field of which consists of two components: a source dipole resulting from a sink and a source flow, and a force dipole corresponding to the rotational flow of a pair of opposite point forces (Fig.~\ref{fig:dipoles}). 

We find that both flow field  components contribute to the reorientation dynamics of the swimmer, albeit in a rather different manner. We demonstrate that the torque exerted by the force dipole, depends crucially on the squirmer characteristics (pusher or puller). On the other hand, the torque arising from the source dipole term (neutral squirmer) has been predicted to be proportional to the viscosity difference~\cite{Simone_SR_2021,Malgaretti_SM_2016}. Our simulations agree with this. 
We show that these two contributions are independent of each other. 

We take into account both the source (Fig.~\ref{fig:dipoles}a) and the force dipole contributions (Fig.~\ref{fig:dipoles}b) as well as the  viscosity ratio $\lambda$, and construct a state diagram for the steady state orientation. We observe that in the steady state, weak swimmers, dominated by the source-dipole contribution, become stationary and orient perpendicular to the interface pointing towards the lower viscosity fluid.  Further, we show that the force dipoles are insensitive to the viscosity contrast but the hydrodynamic boundary condition at interface leads to a reorienting torque. Strong pullers are observed to turn perpendicular to the interface, and become immobile, while strong pushers swim along the interface, pointing towards the lower viscosity fluid with a finite angle respect to the interface normal.

{\it Computational model and parameters.--}

To simulate the finite size squirmers~\cite{Juho_SM_2016,Zaiyi_EPJE_2018}, we impose a slip velocity at the particle surface~\cite{Pagonabarraga_JNFM_2010,magar2003nutrient}
\begin{equation}
 v_{s} = B_{1} \sin\theta + B_{2} \sin\theta \cos\theta ,
\label{eq:slip velocity}
\end{equation}
where $\theta$ is the polar angle with respect to the particle axis $\mathbf{e}$. 
The first term corresponds to a source dipole, which is responsible for the bulk swimming speed $U_0 = \frac{2}{3}B_1$ along the axis $\mathbf{e}$, and generates a far-field varying as $B_1 r^{-3}$ with the distance $r$ from the particle center~\cite{SM}. The second term arises from a force dipole, with a far-field component $\propto B_2 r^{-2}$ that is at the origin of long-range hydrodynamic interactions. The ratio of their amplitudes defines a squirming parameter $\beta=B_2/B_1$, which characterises the swimmer type: pullers (pushers) correspond to $\beta > 0$ ($\beta < 0$) while $\beta = 0$ is a neutral swimmer. 

\begin{figure}[t]
\includegraphics[width=\columnwidth]{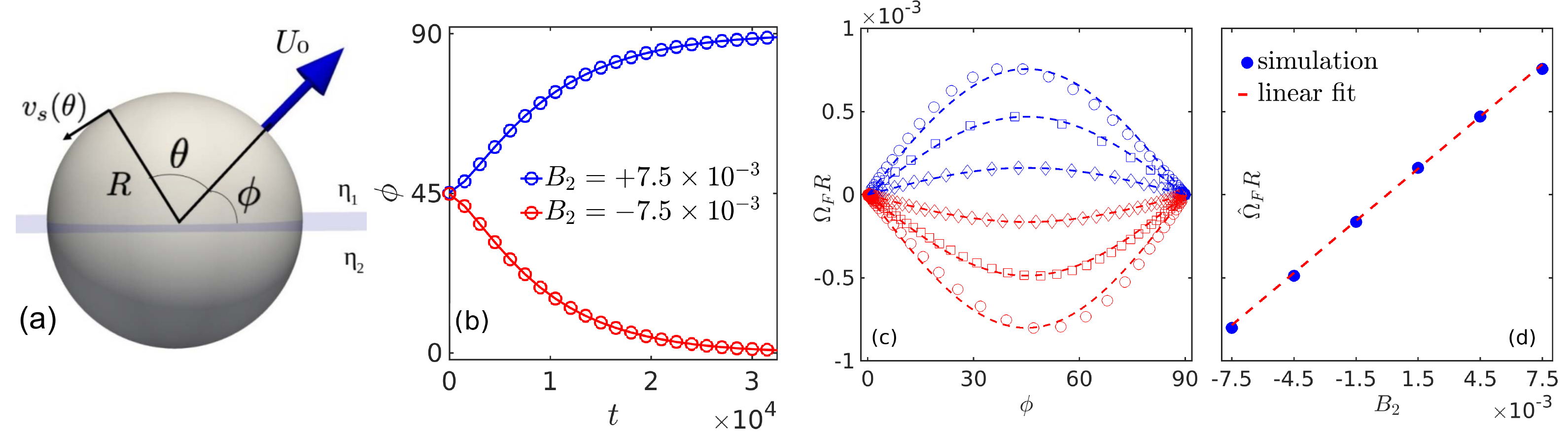}
\caption{(a) Schematic representation of an active squirmer radius $R$ and swimming speed $U_{0}$ at an interface separating two fluids with viscosities $\eta_{1}$ and $\eta_{2}$. The orientation angle $\phi$ is defined as the angle measured between the squirmer orientation and fluid-fluid interface and $\theta$ is the polar angle. (b) Temporal evolution of the orientation $\phi (t)$) for $B_{2}<0$ and $B_{2}>0$ for $\eta_{1}=\eta_{2}$.  
(c) The measured angular velocity $\Omega(\phi)$  as a function of the angle $\phi$.
The blue (red) corresponds to $B_2>0$ ($B_2<0$) and dotted lines are fits to $\hat\Omega_{F}\sin(2\phi)$. (d) The prefactor $\hat\Omega_F$ as a function of the force dipole strength $B_2$.} 
\label{fig:schematic}
\end{figure}

The fluid-fluid interface is realised in terms of a Ginzburg-Landau free energy functional~\cite{Kendon_JFM_2001},
 \begin{equation}\label{eq:free_energy}
 F[c] = \int dV \left({-\frac{A}{2}c^2 + \frac{B}{4}c^4 + \frac{\kappa}{2} |\nabla c|^2} \right)
 \end{equation}
where $-A = B >0$ and $\kappa$ are constants, and $c$ is the phase composition, where $c^*=\pm 1$ are the equilibrium compositions. 
The temporal evolution of the phase field variable $c$ is governed by a Cahn-Hilliard advection-diffusion equation, and the fluid velocity is obtained by solving the incompressible Navier-Stokes equation (for more details of the model see {\it e.g.}~\cite{Kendon_JFM_2001,Hari_PoF_2021}).
The coupled equations are solved using a hybrid finite difference lattice Boltzmann scheme detailed in~\cite{Kendon_JFM_2001,Hari_PoF_2021}.
The phase dependent viscosities are implemented through the relation~\cite{Langaas_EPJB_2000},
\begin{equation}\label{eq:Arrhenius}
\eta (r) = \eta_1^{\frac{1+c}{2}} \eta_2^{\frac{1-c}{2}}  
\end{equation}
where the viscosity takes the values $\eta_{1,2}$ well above or below the interface where $c = \pm1$. {\color{black}This law expresses the fact that the viscosity of liquid mixtures varies exponentially with the concentrations of their components, $\ln\eta=c_1\ln\eta_1+c_2\ln\eta_2$, as first proposed by Arrhenius in 1887 \cite{Arrhenius1887}. In physical terms, it is related to the fact that in many liquids the viscous motion arises from activated jumps, such that $\ln\eta_i$ is a measure of the free enthalpy barrier of molecular component $i$.}

Unless otherwise mentioned we fix $U_0=\frac{2}{3}B_1=10^{-3}$, and vary $B_2$ to study the relative contributions between source and force dipole flows. We consider a particle with radius $R = 12$ in a simulation domain $160\times160\times160$ with periodic boundary conditions. 
The lattice spacing $\Delta x$, time step $\Delta t$ and density $\rho$ are set to unity. The binary fluid parameter are chosen as $B = -A = 0.0258$, surface penalty $\kappa = 0.04$ and mobility $M$ = 0.5. This leads to a flat interface at $c=0$ with an interfacial width $\chi_{0} = \sqrt{2 \kappa/|A|} \approx 1.76$ and interfacial tension $\sigma= \sqrt{8 \kappa |\mathcal{A}|^3/9 \mathcal{B}^2}\approx 0.03$ \cite{Kendon_JFM_2001}. 

The relevant non-dimensional quantities are the capillary number $\mathrm{Ca} = \eta_{1} U_0/\sigma$ which compares viscous stresses with interfacial tension, and the Reynolds number $\mathrm{Re}= \rho R U_{0}/\eta_{1}$ which is the ratio of inertial and viscous forces. With the above parameters, we find $\mathrm{Re}\sim 10^{-2}$ and $\mathrm{Ca} \sim 10^{-2}$, which means that inertial and interfacial deformation effects are negligibly small. 

Using physical parameters of water, and a particle radius $\sim 1\mu$m, $\mathrm{Re}\sim 10^{-2}$ corresponds to $U_0\sim 10^{-2}$m/s. We can map a single length and time unit as $\Delta x\sim 0.1\mu$m, and $\Delta t\sim 10^{-5}$s.

{\it Results.--}  A neutrally wetting particle placed at the interface, adopts a symmetrical position (Fig.~\ref{fig:schematic}a)~\cite{Hari_PoF_2021}. 
We start with the case $B_1 = 0$, where the particle has zero linear velocity yet is subject to a force dipole $B_2$ (Fig.~\ref{fig:dipoles}b). In the bulk, such a ``shaker particle'' does not move yet produces long-range flow field components proportional to $r^{-2}$ and $r^{-4}$ \cite{SM}. When trapped at an interface, however, the particle shows rotational motion in respect to the interface (Fig.~\ref{fig:schematic}). The final stable orientation depends on its squirmer characteristics: A puller ($B_2>0$) turns its axis $\mathbf{e}$ towards the interface normal, and a pusher ($B_2<0$) parallel to the interface (Fig.~\ref{fig:schematic}b). 

We have measured the reorientation velocity $\Omega_F$ resulting from a force dipole, as a function of the angle $\phi$ between particle axis and interface, and find a sinusoidal dependence 
\begin{equation}
    \Omega_F = \hat\Omega_F \sin(2\phi),
    \label{eq:Omega_F}
\end{equation}
as shown by the symbols in Fig.~\ref{fig:schematic}c. The prefactor is proportional to the squirmer parameter, $\hat\Omega_F\propto B_2$ (Fig. ~\ref{fig:schematic}d).

\begin{figure}[tb] 
\includegraphics[width=0.7\columnwidth]{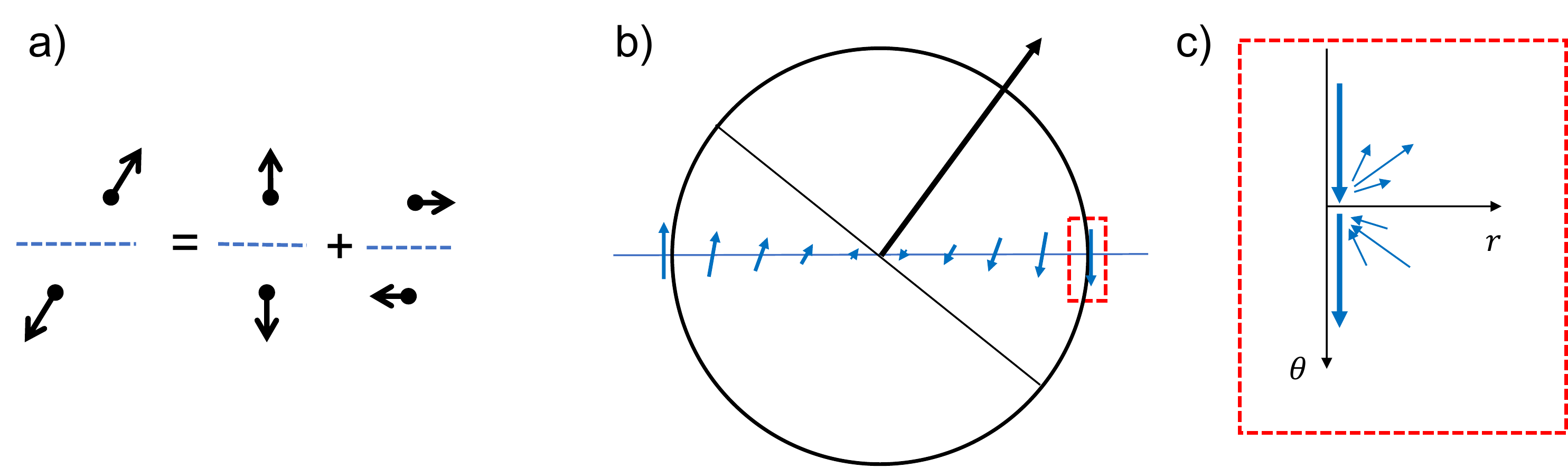}
\caption{a) Decomposition of the force dipole at an angle $\phi$ with respect to a rigid interface with slip boundary conditions. The second term exerts a torque on the particle. The boundary conditions at the particle surface and fluid interface require additional higher multipoles \cite{Blake1971}. b) Slip velocity $v_s$ at the contact line. c) The interface imposes a discontinuity of $v_s$ and an intricate flow profile in the vicinity of the contact line.}
\label{torque} 
\end{figure} 

The source dipole slip-flow is symmetric at the interface (see {\it e.g.} Fig.~\ref{fig:dipoles}a). Thus, 
the rotational motion can only result from the coupling of the 
interface to the force dipole  (Fig.~\ref{fig:dipoles}b), $\mathbf{f_\pm}=\pm\mathbf{e}f_0\delta(\mathbf{r}\mp a\mathbf{e})$, with the particle axis $\mathbf{e}$ and where the squirmer coefficient is defined as $B_2=f_0a/2\pi\eta R^2$. 
In  Fig. \ref{torque}a the force dipole is decomposed in analogy to Blake's treatment at a solid surface \cite{Blake1971}. This gives rise to two terms compatible with a highly rigid fluid interface. From their symmetry it is clear that the first one does not affect the particle motion, whereas the second one results in the torque
\begin{equation}
  \mathbf{T} =4\pi\eta R^2 \, \mathbf{e \cdot n (n \times e)} = 2\pi\eta R^2 \sin(2\phi) \mathbf{\tau},
  \end{equation}
where the unit vector $\mathbf{\tau}$ is perpendicular on the particle axis $\mathbf{e}$ and the interface normal $\mathbf{n}$.

This rationalizes the dependency $\Omega_F\propto \sin 2\phi$ observed in the simulations, yet it does not provide the drag coefficient $T/\Omega_F$, which is determined by the additional contributions to the velocity field. Fig.~\ref{torque}b shows the slip velocity \eqref{eq:slip velocity} at the contact line, which is clearly incompatible with the presence of the interface. Starting from the force-dipole flow and satisfying the boundary conditions both at the interface and at the particle surface, results in a series of source and force multipoles, similar to that of an interfacial particle driven by a self-generated Marangoni flow \cite{Wuerger2014}. The resulting flow profile in the vicinity of the contact line, is illustrated in Fig. \ref{torque}c.

\begin{figure}[tb]
\includegraphics[width=0.5\columnwidth]{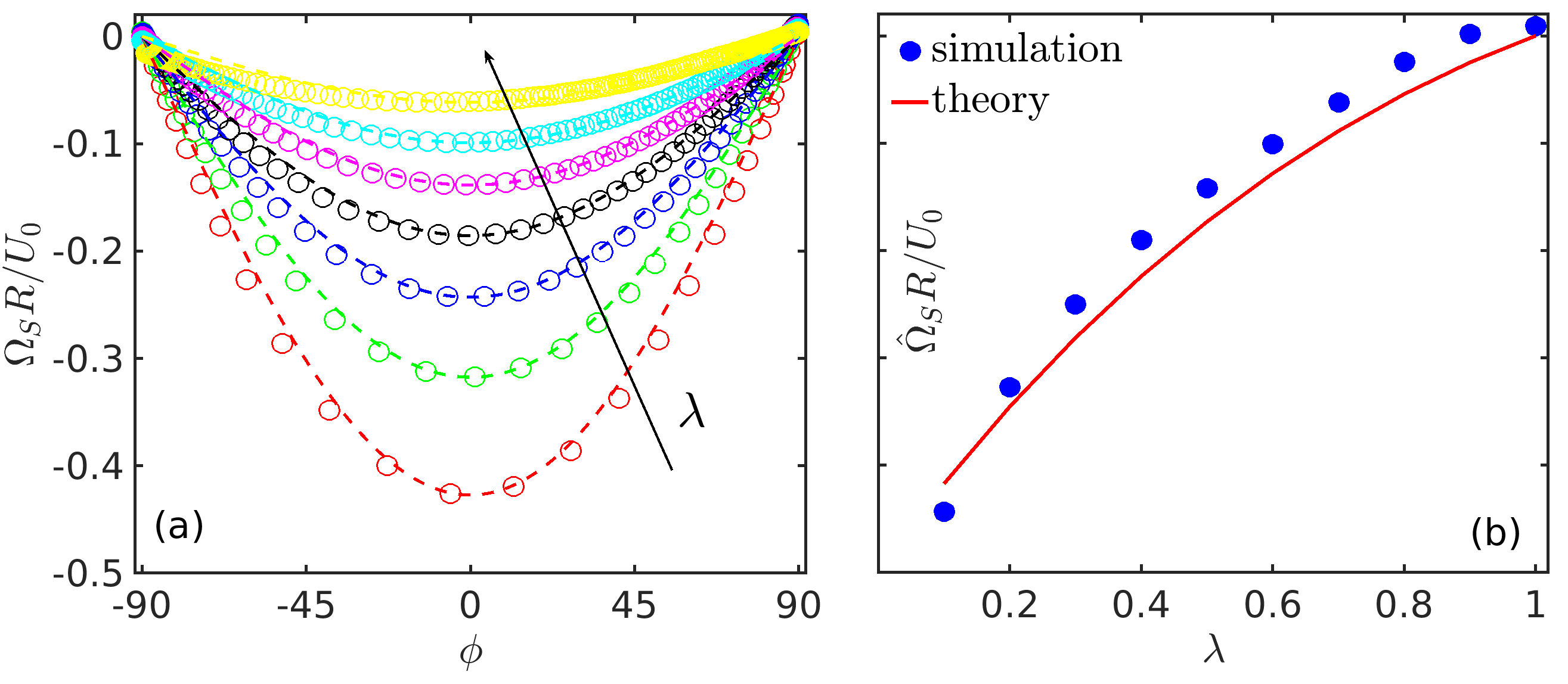}
\caption{ (a) Angular velocity $\Omega_S(\phi)$ observed for a neutral squirmer for different values of the viscosity ratio $\lambda$. The arrow indicates an increase from $\lambda=0.1$ to 0.7, with a step size of 0.1. The dashed lines are given by $\Omega_S=\hat\Omega_S \cos\phi$. (b) The prefactor $\hat\Omega_S$ obtained from  the simulations (full circles), whereas the solid line is calculated from  by eq.~\eqref{eq:omega_theory} (solid line) with the constant $-0.225 U_0/R$.} 
\label{fig:omega_vs_phi_lambda_beta_dependence}
\end{figure}

{\it Source dipole contribution.--} Now we turn to self-propelling microswimmers with a finite $B_1$, which have been shown to be sensitive to variations of the viscosity of the swimming medium. In previous experimental and theoretical investigations,  a negative viscotaxis has been observed~\cite{DANIELS_microbiology_1980,Takabe_microbiology_2017,Liebchen_PRL_2018,Datt_PRL_2020,Eastham_PRF_2020,Simone_SR_2021,Stehnach_Biorxiv_2020,lopez_arxiv_2020,Malgaretti_SM_2016}. To study the effect of the source dipole $B_1$ we consider a neutral squirmer ($B_2=0$) and introduce  a viscosity ratio $\lambda=\eta_2/\eta_1$. 

Our simulation data for the source-dipole driven angular velocity $\Omega_S$ are shown in Fig.~\ref{fig:omega_vs_phi_lambda_beta_dependence}, as a function of the particle orientation $\phi$ and the viscosity ratio $\lambda$. For $\lambda\leq0.7$ the variation with the angle $\phi$ (Fig.~\ref{fig:omega_vs_phi_lambda_beta_dependence}a) obeys the relation 
\begin{equation}
\Omega_S=\hat\Omega_S\cos\phi,
\end{equation}
as expected from previous work \cite{Datt_PRL_2020,Malgaretti_SM_2016}. For larger values of $\lambda$, the hydrodynamic torques are weak and a slow  drift of $\phi(t)$ is observed, which likely arises from numerical errors due to the finite capillary number used in the simulations. In Fig.~\ref{fig:omega_vs_phi_lambda_beta_dependence}b we plot the dependence of the prefactor $\hat\Omega_S$ on the viscosity ratio $\lambda$, and find a good agreement with the theoretical prediction \cite{Malgaretti_SM_2016}
\begin{equation}
\label{eq:omega_theory}
\hat\Omega_S= \mathrm{const.}\times\frac{(2-\lambda)(1-\lambda)}{1-\frac{3}{32}(1-\lambda)^2}
\end{equation}
with the constant $-0.225 U_0/R$.  

\begin{figure}[tb]
\includegraphics[width=0.5\columnwidth]{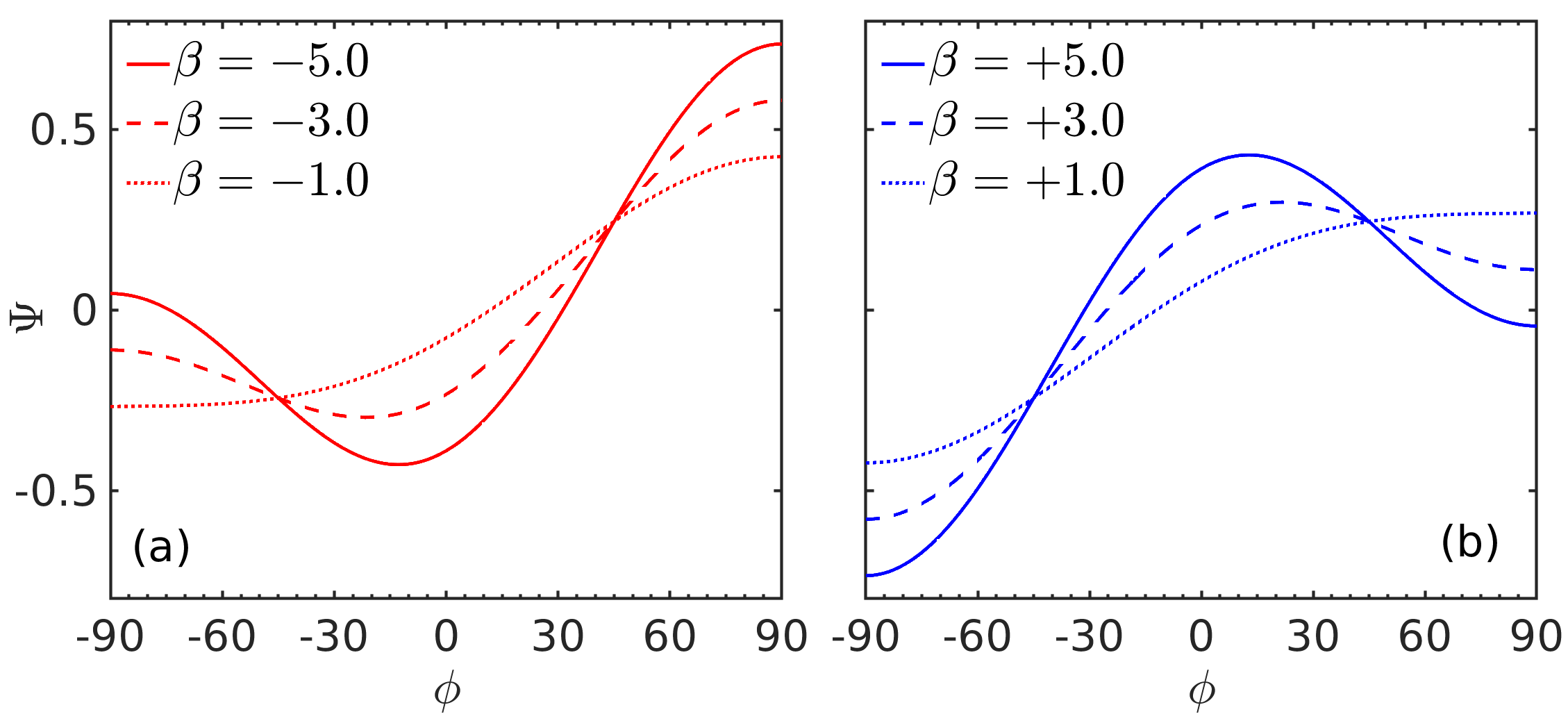}
\caption{Orientational potential $\Psi(\phi) = - \int_0^\phi{\Omega(\phi') d\phi'}$ for pushers ($\beta<0$) and pullers ($\beta>0$) with $\lambda = 0.2$, calculated from eq. \eqref{eq:omega_total}. }
\label{fig:lambda_beta_contribution_fixed_points}
\end{figure}

{\it General squirmer.--} Finally, we consider the general case of a microswimmer with both force and source dipoles. At low Reynolds number the reorientation rate  can be written as the sum of their respective contributions, $\Omega=\Omega_S+\Omega_F$, where the former is proportional to $B_1$ and the latter to $B_2$. In the following we assume a constant self-propulsion velocity $U_0=\frac{2}{3}B_1$, and discuss the dependencies on the viscosity ratio $\lambda$ and the squirmer parameter $\beta=B_2/B_1$, 
\begin{equation}
\label{eq:omega_total}
\Omega(\lambda,\beta) = \hat\Omega_S(\lambda)\cos\phi + \hat\Omega_F(\beta)\sin2\phi. 
\end{equation}
The prefactor $\hat\Omega_S$ is a  complex function of $\lambda$, whereas $\hat\Omega_F$ is proportional to $\beta$. The stationary points $\phi^*(\lambda,\beta)$ of the orientational dynamics are given by a zero angular velocity, $\Omega = 0$, and can be visualized by an effective orientational potential $\Psi = -\int{\Omega d\phi}$ (Fig.~\ref{fig:lambda_beta_contribution_fixed_points}). From eq. \eqref{eq:omega_total} it is clear that its minima and maxima are given by $\phi^* =\arcsin(\hat\Omega_S/2\hat\Omega_F)$ for $|\hat\Omega_S| < |2\hat\Omega_F|$, corresponding to strong pushers and pullers, respectively. For weak force dipoles, $|\hat\Omega_S| \geq |2\hat\Omega_F|$, only two stationary points $\phi^* = \pm 90^\circ$ are observed (Fig.~\ref{fig:lambda_beta_contribution_fixed_points}).
  
\begin{figure}[tb]
\includegraphics[width=0.5\columnwidth]{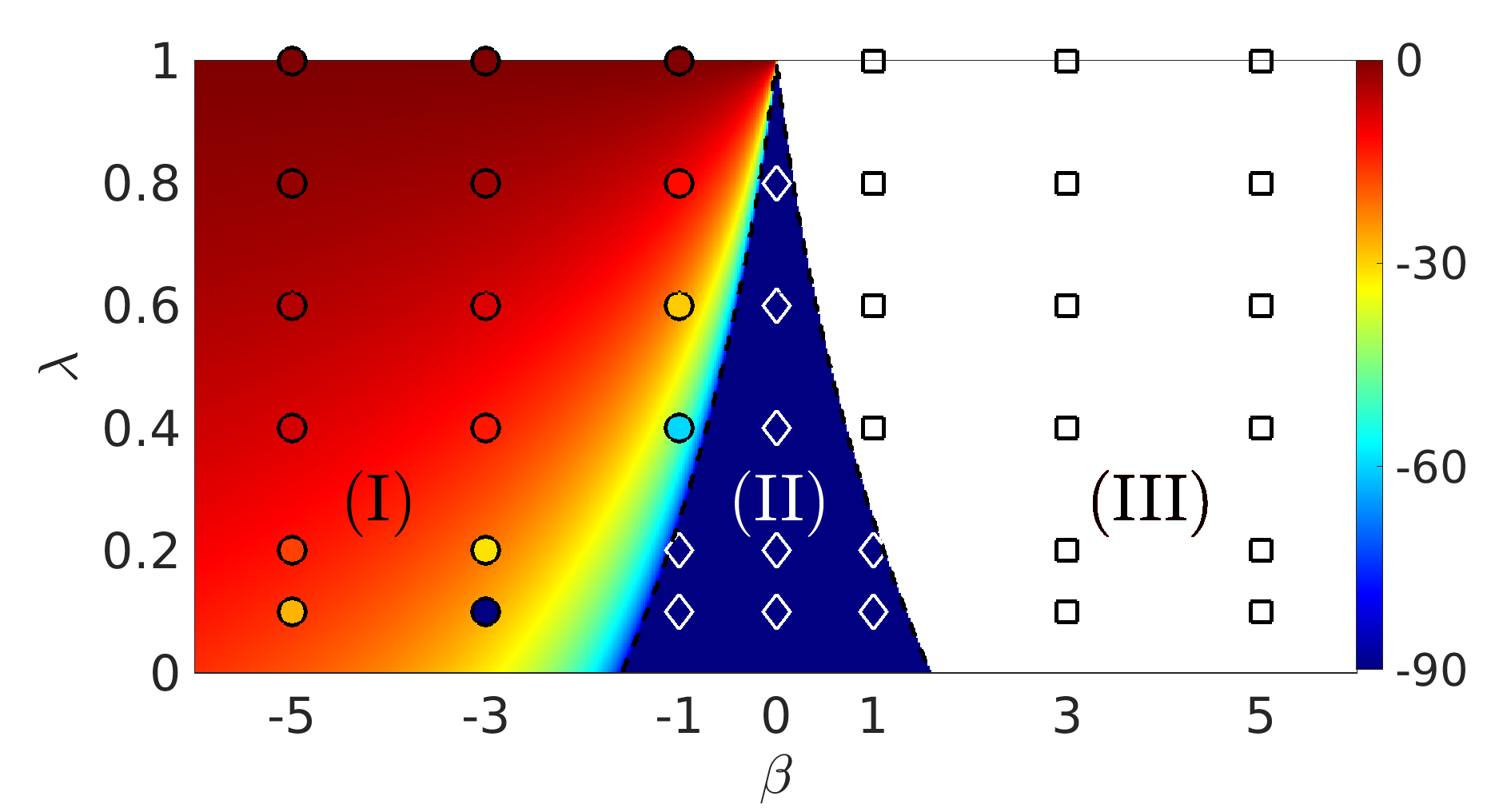}
\caption{A state diagram for the stationary angle $\phi^*$ in a $\beta$ - $\lambda$ space. Simulation results (symbols) and theory (background) are color-coded according to the color bar at the right. Range I (circles) indicates a stable fixed point in the range $-90^\circ < \phi \leq 0$ for pushers, and range II (diamonds) at $-90^\circ$ for sufficiently weak pusher and pullers, where the source dipole contribution dominates. 
Range III (squares) indicates the stationary states $\phi^*=\pm90^\circ$ for strong pullers.  The dashed lines give the boundaries between these states and are calculated from $\hat\Omega_S=\pm2\hat\Omega_F$. (See text for more details)} 
\label{fig:phase_diagram}
\end{figure}

Using \eqref{eq:omega_total}, we construct a steady state diagram in terms of stable and unstable fixed point in the $\lambda-\beta$-space (Fig.~\ref{fig:phase_diagram}). 
To test these theoretical predictions, we carried out simulations where the squirmer parameter $\beta$ and the viscosity ratio $\lambda$ were varied. Both theory and simulations show three different ranges. 

The first one (I) is observed for pushers ($\beta < 0$)  and is characterized by a stable fixed point between $-90^\circ$ and $0^\circ$ (Fig.~\ref{fig:phase_diagram}), corresponding to minimum of $\Psi$ (see {\it e.g.} $\beta =-3$ and $\beta = -5$ curves in Fig.~\ref{fig:lambda_beta_contribution_fixed_points}). This arises from the competition between the force dipole contribution turning the particle towards the interface and the source dipole, which turns the particle towards the lower viscosity fluid. The resulting steady-state orientation varies from parallel to the interface ($0^\circ$, upper left corner) to normal orientation ($-90^\circ$, dashed line). The $\phi^*$ observed from the simulations are given by circles, with a filling color according to the color bar at the right in Fig.~\ref{fig:phase_diagram}. The background color corresponds to the theoretical expression $\phi^* =\arcsin(\hat\Omega_S/2\hat\Omega_F)$, where the constant prefactors of $\hat\Omega_F$ and $\hat\Omega_S$ 
are taken from the fits in Figs. \ref{fig:schematic}d and \ref{fig:omega_vs_phi_lambda_beta_dependence}b, respectively.   

In range (II) the stationary orientation corresponds to a minimum at $\phi=-90^\circ$, which occurs for  sufficiently small $|\beta|$, where the source dipole term $\Omega_S$ is dominant and orients the swimmers orient towards the lower viscosity fluid. The dashed lines give the theoretical boundaries of range II, defined by $\Omega_S=\pm 2\Omega_F$. The upward orientation  $\phi=90^\circ$ corresponds to an unstable fixed point (see {\it e.g.} $\beta=\pm1$ curves in Fig.~\ref{fig:lambda_beta_contribution_fixed_points}). 

Range (III) describes strong pullers. The force dipole contribution $\Omega_F$ dominates and turns the swimmer towards one of the minima at $\phi=\pm90^\circ$. The steady state orientation is decided depending on whether the initial orientation is below or above the unstable fixed point at $\phi=\arcsin(\hat\Omega_S/2\hat\Omega_F)$, as illustrated by the curves for $\beta=+3$ and +5 in Fig.~\ref{fig:lambda_beta_contribution_fixed_points}. 

We observe a good agreement between the simulations and the theory based on the superposition principle of the two interactions $\eqref{eq:omega_total}$. The deviations observed for $\beta < 0$ at strong viscosity contrast, $\lambda<0.2$, are possibly due to numerical artefacts overestimating the source dipole contribution.

{\it Conclusions.--}
We have investigated the reorientation dynamics of spherical microswimmers trapped at a clean fluid-fluid interface. In rather good agreement with the theoretical models, our numerical simulations demonstrate that the reorientation has two, independent, components: The force dipoles give a rise to a torque, which drives a parallel steady-state orientation for pushers and a perpendicular one for pullers. When a viscosity difference is introduced, our simulations show that neutral swimmers orient towards the lower viscosity fluid, in agreement with simulations of catalytic particles~\cite{Malgaretti_SM_2016} and bacterial experiments~\cite{Simone_SR_2021, Stehnach_Biorxiv_2020} in sharp viscosity profiles. Our results show moreover, that these two contributions are independent of each other, and that their interplay defines the steady-state orientation. 

In summary,  in the case of a weak force dipole and strong viscosity contrast, the swimmers align on the interface normal toward the less viscous fluid, such that their translational velocity is zero. Similarly, strong pullers adopt a normal orientation yet may be trapped in an upward or downward state and become stationary. A different behavior is predicted for strong pushers, which in the steady state reach a finite inclination angle,  toward the lower-viscosity fluid, and thus move at a finite velocity $U= U_0\cos\phi^*$ along the interface.

{\it Acknowledgements.--}
HG would like to thank Sotiris Samatas for engaging and helpful discussions. HG is grateful for the computational resources provided at LOMA (University of Bordeaux) and cluster Curta at MCIA. HG, ZS and JSL acknowledge the French National Research
Agency (ANR) through Contract No. ANR-19-CE06-0012-01, IdEx Bordeaux and la r\'egion Nouvelle-Aquitaine for funding.

\nocite{*}

\bibliography{apssamp}

\end{document}